\documentclass[cits]{PoS}
\title{Simulations of Cold Electroweak Baryogenesis\footnote{Submitted 23-10-2005}}

\author{\speaker{Anders Tranberg}\\
University of Sussex,\\
E-mail: \email{a.tranberg@sussex.ac.uk}}
        
\author{Jan Smit,\\
University of Amsterdam,\\
E-mail: \email{jsmit@science.uva.nl}}

\abstract{Cold Electroweak Baryogenesis is an attempt to explain the cosmological baryon asymmetry using only a minimal extension of the Standard Model. The relevant processes take place out of thermal equilibrium and are non-perturbative, and so must be studied using full, real-time lattice simulations. We present new results on the dependence on CP-violation and the Higgs-to-W-mass ratio.}

\FullConference{XXIIIrd International Symposium on Lattice Field Theory\\25-30 July 2005\\ Trinity College, Dublin, Ireland}

\ShortTitle{Simulations of Cold Electroweak Baryogenesis}

\PoS{PoS(LAT2005)246}

\newcommand{\Tr}{\mbox{Tr}\,}  

\begin{document}


\section{Cold Electroweak Baryogenesis}
In the Standard Model, C and CP are broken symmetries and baryon number is violated at high temperatures \cite{Rubakov:1996vz}. In the early evolution of the Universe, reheating after inflation and subsequent cooling through expansion supplied out-of-equilibrium conditions, making it possible for a baryon asymmetry to be generated. Cold Electroweak Baryogenesis attempts to achieve this using only Standard Model physics, minimally extended with one scalar inflaton field.

Inflation is assumed to end at the electroweak scale ($\simeq 100$GeV), after which the rolling of the inflaton triggers electroweak symmetry breaking through a change of sign in the effective Higgs potential,
\begin{eqnarray}
V(\sigma,\phi)=V(\sigma) + \left(\mu^{2}-\lambda_{\sigma\phi}\sigma^{2}\right)\phi^{\dagger}\phi + \lambda(\phi^{\dagger}\phi)^{2}+V_{0}.
\end{eqnarray}
The inflaton potential should provide inflation \cite{vanTent:2004rc}, and when $\mu_{\rm eff}^{2}(t) = \mu^{2}-\lambda_{\sigma\phi}\sigma^{2}$ goes through zero, Higgs symmetry breaking is triggered, leading to tachyonic preheating \cite{Felder:2000hj,Skullerud:2003ki}. This out of equilibrium transition will supply the required conditions for baryogenesis to take place \cite{Krauss:1999ng,Garcia-Bellido:1999sv,Copeland:2001qw,Tranberg:2003gi,Garcia-Bellido:2003wd}. 

Baryon number $B$ is violated in the SM through a quantum anomaly, relating the change in $B$ to the change in Chern-Simons number $N_{\rm cs}$ of the $SU(2)$ gauge fields, to which the fermions are (axially) coupled,
\begin{equation}
B(t)-B(0)= 3\langle N_{\rm cs}(t)-N_{\rm cs}(0)\rangle= 3 \int_{0}^{t} dt\,d^{3}x \langle\dot{n}_{\rm cs}\rangle=3 \int_{0}^{t} dt\,d^{3}x \langle\frac{1}{16\pi^{2}}\Tr F^{\mu\nu}\tilde{F}_{\mu\nu}\rangle.
\label{anomaly}
\end{equation}
We will focus on the part of the Standard Model important for the baryon number violating processes, namely the $SU(2)$-Higgs system. In other words, we will neglect the dynamics of the fermions, $U(1)$ and $SU(3)$ gauge fields and the inflaton. In particular, we will assume that the inflaton is simply responsible for an instantaneous quench,
\begin{equation}
\mu^{2}_{\rm eff}(t) = \mu^{2},~~~ t<0,\qquad \mu_{\rm eff}^{2}(t) = - \mu^{2},~~~ t>0.
\end{equation}
In order to generate an asymmetry, we need to introduce CP-violation explicitly in the model. We do this by including the lowest order CP-violating term composed of gauge and Higgs fields,
\begin{equation}
\mathcal{L}_{\rm CP}=\kappa\phi^{\dagger}\phi \Tr F^{\mu\nu}\tilde{F}_{\mu\nu}.
\label{cpterm}
\end{equation}
It is easy to see, that this amounts to a chemical potential for Chern-Simons number density $n_{\rm cs}$ (eq. \ref{anomaly}), thus biasing the baryon number.

The generation of the baryon asymmetry is intrinsically non-perturbative, non-linear and out of equilibrium. We therefore need to numerically simulate the full, real-time dynamics on the lattice. We do this by applying the classical approximation. Apart from being practically convenient, this approximation is well motivated by the (exponentially) large occupation number in the Higgs and gauge fields during and after the transition \cite{Garcia-Bellido:2002aj,Smit:2002yg,Skullerud:2003ki}. In particular, since the highly occupied low momentum modes are also the ones relevant for the baryon number violating processes, we expect those to be well reproduced by the classical equations of motion.


\section{Numerical setup}
The continuum action of the $SU(2)$-Higgs:
\begin{equation}
S=-\int dt d^{3}x\,\left[\frac{1}{2g^{2}}\Tr F^{\mu\nu}F_{\mu\nu}+(D_{\mu}\phi)^{\dagger}D^{\mu}\phi+\lambda\left(\phi^{\dagger}\phi\right)^{2}+\left(\mu^{2}-\lambda_{\sigma\phi}\sigma^{2}\right)\phi^{\dagger}\phi+V_{0}+\mathcal{L}_{\rm CP}\right].
\end{equation}
is discretized on the lattice and the equations of motion are derived. Discretizing the $F\tilde{F}$ term is notoriously difficult, sine it suffers from UV lattice artefacts. By symmetrizing all the plaquettes and because only the low-momentum modes are highly populated, this problem can be resolved. It does mean that the equations of motion become complicated and, what is worse, implicit, leading to a factor of about 10 in CPU time. 

The only free parameters are the Higgs self-coupling through the Higgs mass $m_{H}^{2}/m_{W}^{2}=8\lambda/g^{2}$ and the CP-violation, parameterized through $\delta_{\rm CP}=16\pi^{2}\kappa m_{W}^{2}/3$. We use a mass of $am_{H}=0.3$ on lattices of size $(Lm_{H})^{3}=(N am_{H})^{3}=27^{3}$.

\begin{figure}
\includegraphics[width=1.0\textwidth]{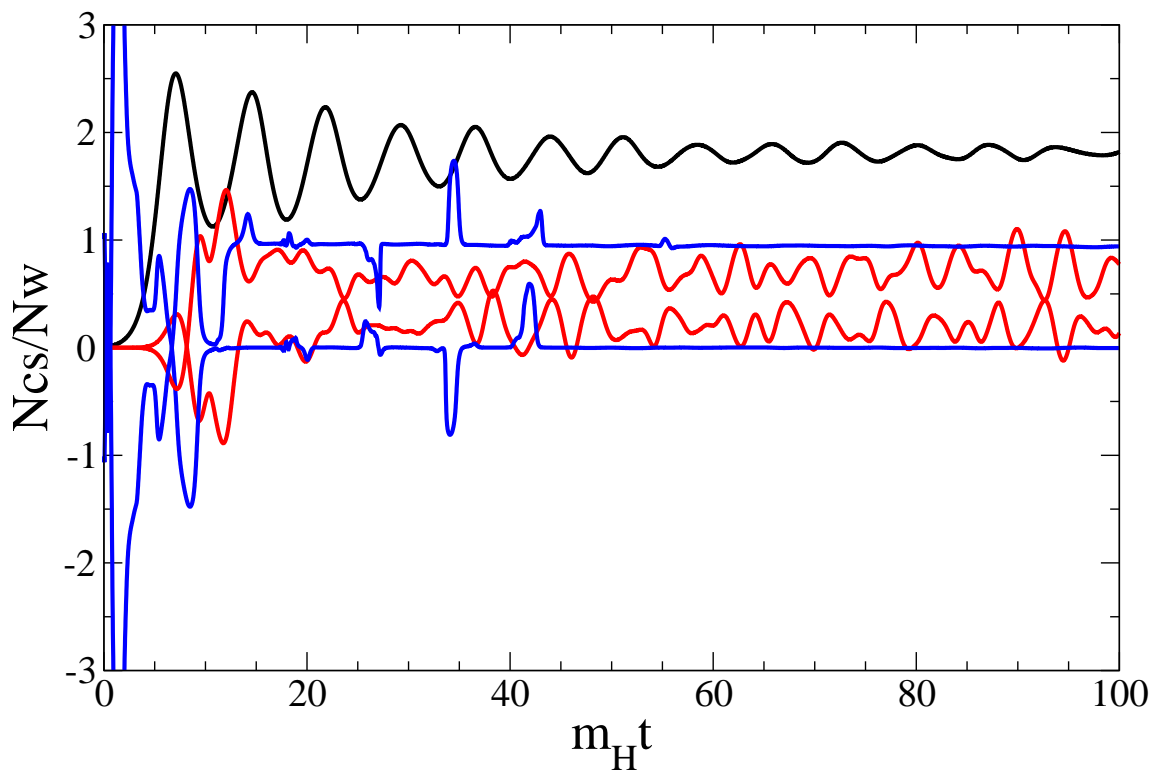}
\label{singletraj}
\caption{Two trajectories from a pair of CP-conjugate initial configurations. $\delta_{\rm CP}=1$, $m_{H}=2m_{W}$. The Higgs field (black) ``rolls off the hill'' and oscillates around the broken phase minimum. Chern-Simons number (red) grows and settles to 0 and 1, respectively, and similarly for the winding number (blue). This amount to a net asymmetry of 1 in Chern-Simons number, 3 in baryon number.}
\end{figure}

The classical equations of motion are solved for a CP-symmetric ensemble of initial conditions, and the evolution of the average Higgs field $\phi^{\dagger}\phi$, Chern-Simons $N_{\rm cs}$ and Higgs winding number $N_{W}$ is studied. An example of two runs starting from a pair of initially CP-conjugate configurations is shown in figure \ref{singletraj}. We choose to concentrate on the Higgs winding number, since it is very closely integer, and so provides a cleaner signal than the Chern-Simons number itself. For late times, $N_{\rm cs}\simeq N_{W}$ in order to minimize the energy contribution from the covariant derivative. Including CP-violation, CP-conjugate configurations indeed end up with different (integer) values of the final winding number. 
%


\section{Dependence on CP-violation}
\begin{figure}
\includegraphics[width=1.0\textwidth]{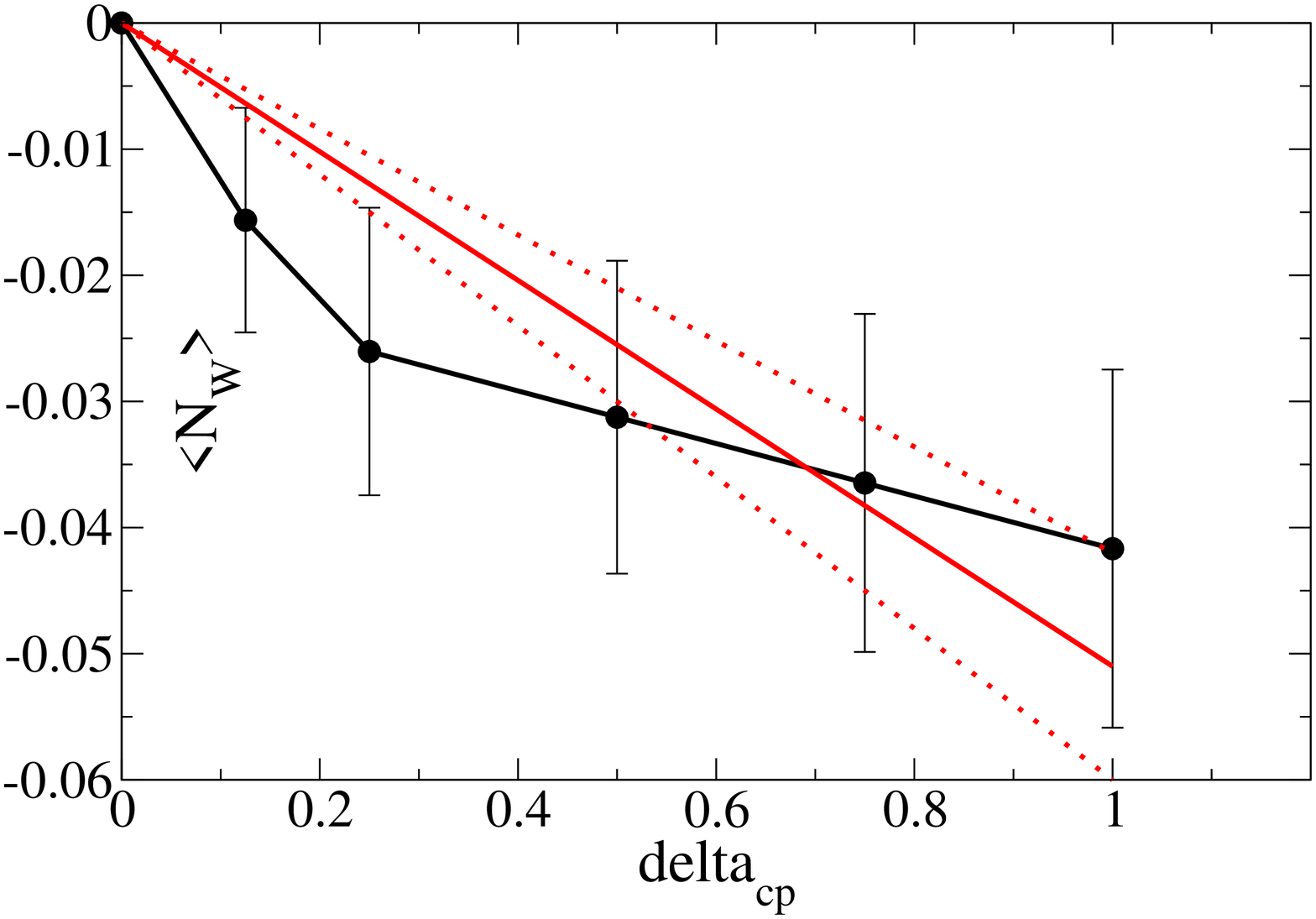}
\label{cpdep}
\caption{$\langle N_{W}\rangle$ vs. $\delta_{\rm cp}$ for $m_{H}=2m_{W}$. The full line is a fit, the two dotted line represent $\pm 1 \sigma$ in the fitted slope.}
\end{figure}

We average over an ensemble of 96 pairs of configurations for a range of values of $\delta_{\rm cp}$. Figure \ref{cpdep} shows an example of $\langle N_{W}\rangle$ vs. $\delta_{\rm CP}$. A one-parameter fit is overlaid (we know that the asymmetry is zero for zero $\delta_{\rm CP}$).   
Given $\langle N_{W}\rangle$, we can write ($g_{*}=86.25$)
\begin{eqnarray}
\frac{n_{B}}{n_{\gamma}}=\frac{7.04}{s}\frac{3\langle N_{W}\rangle}{V},\qquad s=\frac{2\pi^{2}}{45}g_{*}T^{3},\qquad\frac{\pi^{2}}{30}g_{*}T^{4}=V_{0}=\frac{m_{H}^{4}}{16\lambda},
\end{eqnarray}
\begin{eqnarray}
\frac{n_{B}}{n_{\gamma}}&=3.2\times10^{-4}\times\langle N_{W}\rangle\times\left(\frac{m_{H}}{m_{W}}\right)^{3/2}&,\\
\frac{n_{B}}{n_{\gamma}}&=-(0.46\pm 0.08)\times 10^{-4}\times \delta_{\rm CP},&~~m_{H}=2m_{W}.
\end{eqnarray}
%

\section{Dependence on the Higgs mass}
\begin{figure}
\includegraphics[width=0.5\textwidth]{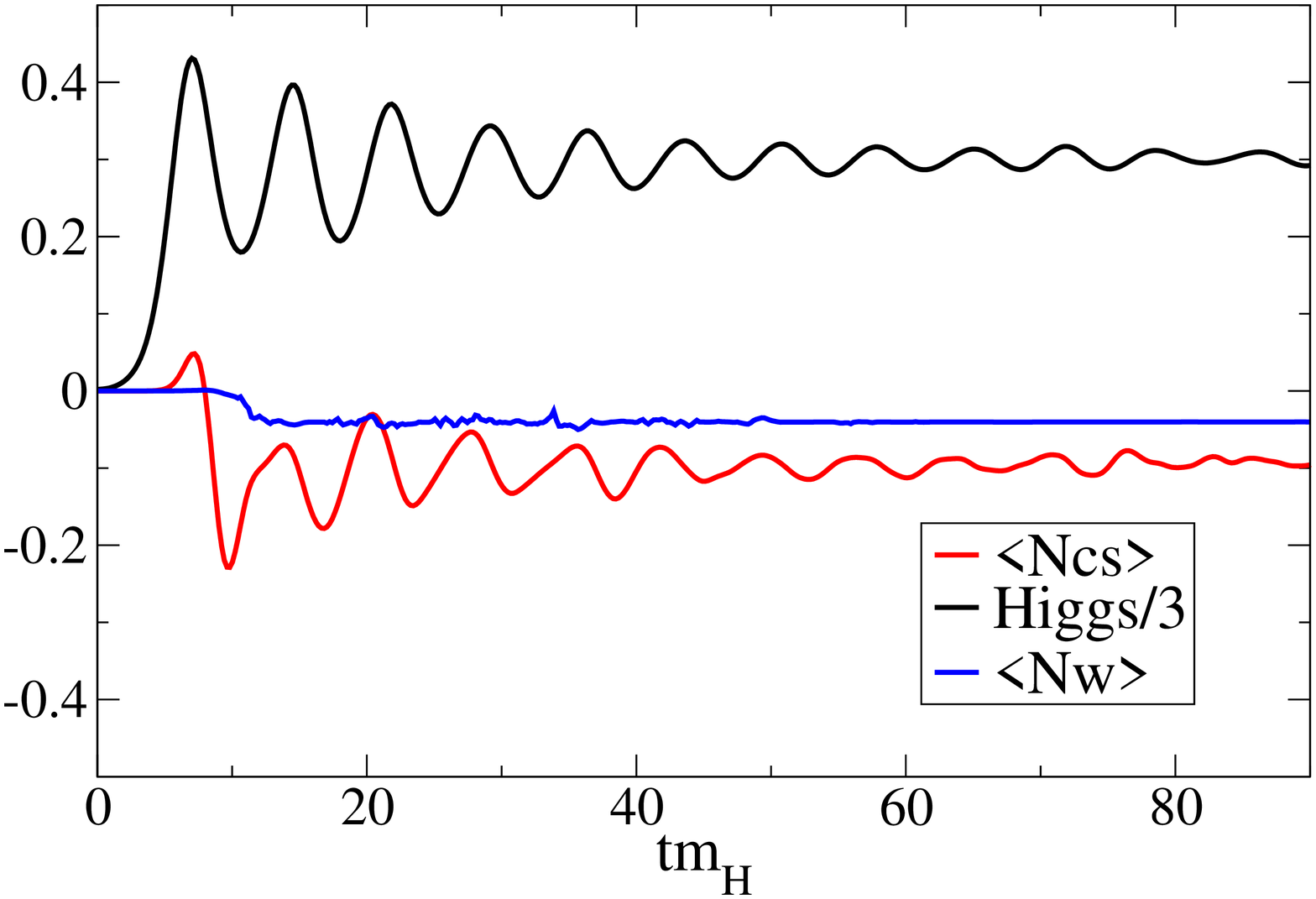}
\includegraphics[width=0.5\textwidth]{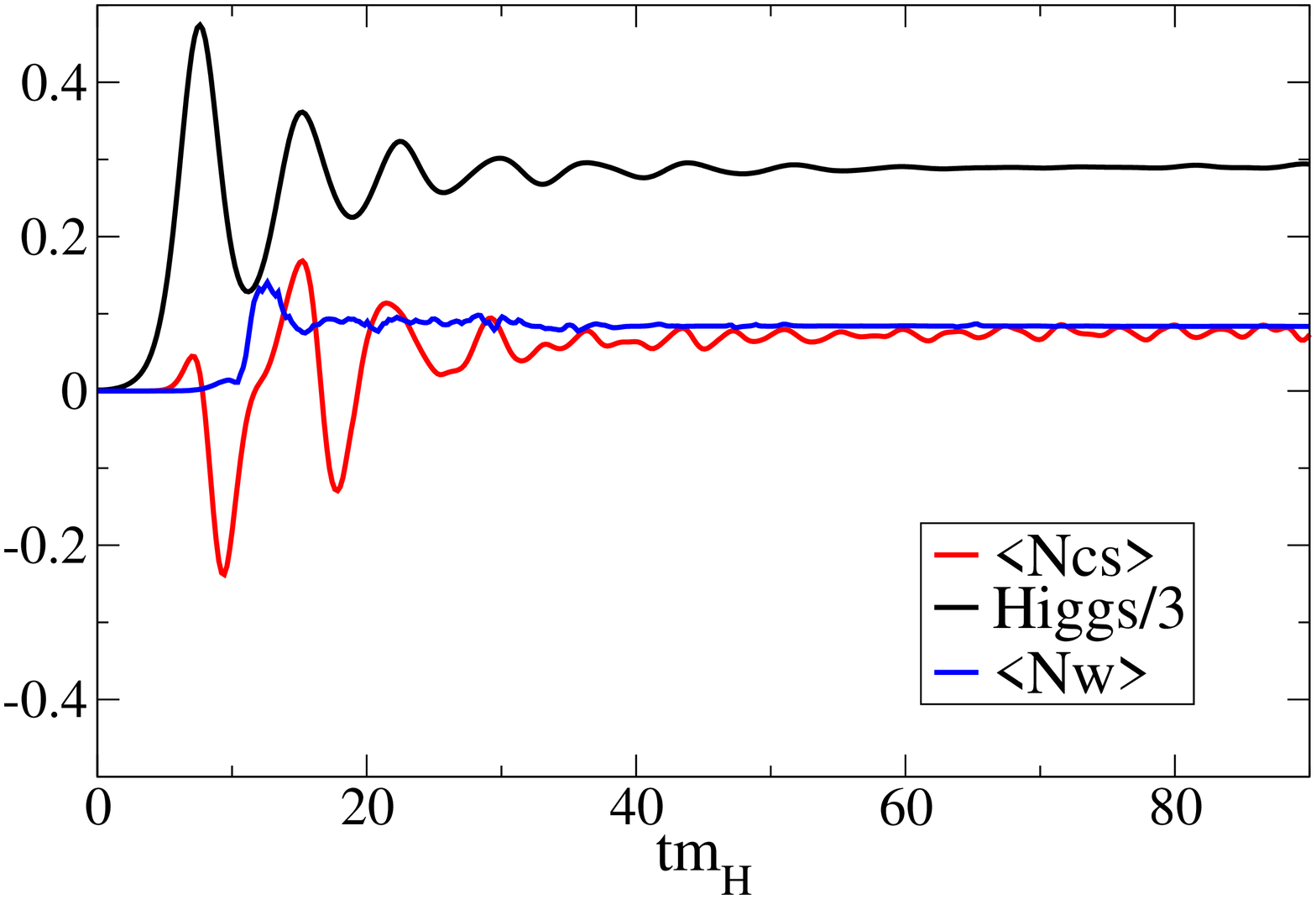}
\label{mdep}
\caption{The time evolution of $\langle\phi^{\dagger}\phi\rangle$ (black), $\langle N_{\rm cs}\rangle$ (red) and $\langle N_{W}\rangle$ (blue) for $m_{H}=2m_{W}$ (left) and $=\sqrt{2}m_{W}$ (right), $\delta_{\rm cp}=1$.}
\end{figure}

In an analogous model in 1+1 dimensions, we found that the baryon asymmetry was very sensitive to the value of the Higgs-to-W-mass ratio \cite{Smit:2002yg}. This was a result of a conspiracy of oscillation frequencies. It makes sense to see whether the present model has a similar feature. Figure \ref{mdep} shows the time evolution of the ensemble averaged $\phi^{\dagger}\phi$, $N_{\rm cs}$ and $N_{W}$ for $m_{H}=2m_{W}$ (left) and $\sqrt{2}m_{W}$ (right). Obviously, in the latter case the final asymmetry has the opposite sign(!) 
A closer look at the evolution reveals what is going on. In the beginning, exponentially growing gauge fields result in exponentially increasing average Chern-Simons number, proportional to $\delta_{\rm CP}$. Around time $m_{H}t=5$, there is change of direction, so that the asymmetry is now proportional to $-\delta_{\rm CP}$. This behaviour can be reproduced along the lines of \cite{Garcia-Bellido:2003wd} as a reaction to a time-dependent chemical potential induced by the CP-violating term (eq. \ref{cpterm}). Only at time $m_{H}t\simeq 10$ is an asymmetry created in the winding number, which at this time changes to accomodate the non-zero Chern-Simons number, after which it moves no more.

The Higgs winding number can only change around a zero of the Higgs field. This is why everything happens around the first minimum of the average Higgs around time $m_{H}t=10$, where there are many such zeros. The winding can then catch up with the Chern-Simons number. It is also around this minimum that the different results for the two different mass ratios is created. Presumably, this has to do with whether the time derivative of the Higgs field (the effective chemical potential) has one sign or the other at that precise time. As in 1+1 dimensions, this is a question of a conspiracy between the Higgs and gauge oscillation frequencies.


\section{Outlook}
We have seen that an asymmetry is indeed generated in the electroweak tachyonic transition in the presence of CP-violation. This was already established in \cite{Tranberg:2003gi}, of which the present work is a continuation. The dependence on the CP-violation ($\delta_{\rm CP}$) is consistent with linear, giving an estimate for the resulting asymmetry. The mass dependence is complicated and reminiscent of the case in 1+1 dimensions, and even the overall sign can depend on it. The fact that the mass dependence seems to have to do with a coincidence of frequencies, suggests that for the physical case of a finite time symmetry breaking transition, such an effect may not be present. This will be discussed in an upcoming publication.


\section*{Acknowledgements}
This work is supported by FOM/NWO and the PPARC SPG ``Classical Lattice Field Theory''.

\end{document}